\documentclass[aps,prl,twocolumn,groupedaddress,latterpaper]{revtex4}
\usepackage{graphicx}

\begin{document}

\title{Manifestation of photonic band structure in small clusters of spherical particles}
\author{Alexey Yamilov}
\email{a-yamilov@northwestern.edu}
\author{Hui Cao}
\affiliation{Department of Physics and Astronomy, Northwestern University, Evanston, IL 60208}
\date{\today}

\begin{abstract}

We study the formation of the photonic band structure in small clusters of dielectric spheres. The first signs of the band structure, an attribute of an infinite crystal, can appear for clusters of 5 particles. Density of resonant states of a cluster of 32 spheres may exhibit a well defined structure similar to the density of electromagnetic states of the infinite photonic crystal. The resonant mode structure of finite-size aggregates is shown to be insensitive to random displacements of particles off the perfect lattice positions as large as  half-radius of the particle. The results were obtained by an efficient numerical method, which  relates the density of resonant states to the the scattering coefficients of the electromagnetic scattering problem. Generalized multisphere Mie (GMM) solution was used to obtain scattering matrix elements. These results are important to miniature photonic crystal design as well as understanding of light localization in dense random media. 

\end{abstract}


\maketitle

The idea of employing photonic crystals, structures with periodically modulated dielectric constant, to manipulate the density of states (DOS) of electromagnetic waves (EM) \cite{yablonovitch,john} led to an explosion of interest, both academic and practical \cite{johnson,vlasov}. In a photonic crystal, in a given crystallographic direction, light can propagate only for the frequencies within photonic bands, described by a dispersion relation $\omega=\omega (\textbf{k})$, similar to electron de Broglie waves in crystals. If the gaps between the bands overlap for all directions in the crystal, photonic crystal posses a complete photonic band gap \cite{ho_soukoulis}. For frequencies inside the complete gap, density of EM states turns to zero, that leads to new phenomena such as suppression of spontaneous emission
, light localization, zero-threshold lasers, all-optical transistors and circuitry, and anomalous nonlinear properties\cite{yablonovitch,john,johnson,vlasov,soukoulis}. For theoretical considerations, the size of the crystal is usually assumed infinite. In practice, however, one always deals with finite structures. Naturally, a question arises: \textit{How large a photonic crystal should be, in order to exhibit photonic band structure?} In this letter, we show that the modification of the density of EM states may begin for a cluster of 5 spheres. This suggests that scattering of light by ordered aggregates of spheres can be important in the problem of light propagation in dense random media. Indeed, an ensemble of mono-disperse, resonant scatterers (e.g. spheres) is considered the most favorable \cite{bart} for satisfying Ioffe-Regel criterion for Anderson localization $kl_{scat}\alt 1$ \cite{ping_sheng,bart,soukoulis,andrey,andrey_nature,wiersma_nature}, where $k$ is the wavenumber in the medium and $l_{scat}$ is the scattering length. Closely-packed face centered cubic (FCC) arrangement often appear in self-assembled structures and it is shown to have lowest free energy. This fact points to a high probability of formation of ordered clusters in a collection of spheres. Even though the complete photonic gap can never be realized in an FCC structure of dielectric spheres in air, for some parameters, the total density of EM states can be significantly suppressed in certain frequency regions.  Our results suggest that such suppression can occur due to the presence of the clusters even without long range order. 

Treatment of photonic structures, with all 3 dimensions of the order of the wavelength has been a challenge. When the distance between scatterers is comparable to their size, analytical methods, such as single scattering approximation, dipole approximation become invalid\cite{bart}. Numerical methods tailored for calculation of transmission through a slab \cite{stacking_faults,pendry} cannot be applied because it requires the lateral size of the slab to be larger than all characteristic lengths. Finite difference time domain (FDTD) method is used extensively in photonics design \cite{taflove,scherer}. Obtaining the density of states with FDTD-based ONYX method \cite{pendry} requires time-dependent solution.  Combined with dense $\lambda/20$ spacial grids, it  makes the problem computationally demanding in 3D, even for the smallest structure considered in this paper. However, the tendency to miniaturization of photonic devices calls for an efficient tool for dealing with such structures. An efficient alternative approach is needed.

For small truncated photonic crystal structures, \textit{scattering language} becomes appropriate. We use the GMM solution \cite{gmm} to obtain the elements of the scattering matrix of the collection of dielectric spheres. The density of resonant EM states of the cluster can be easily expressed in terms of the scattering matrix. Within this approach,  one only needs to evaluate vector-spherical-function expansion coefficients, there is no need, as in FDTD method, to find EM fields at every spacial point. The method requires considerably less memory and computation time resources, that makes it possible to use on a personal computer for up to 300 particles. An important class of photonic crystals, opals, is treated in this framework as an example. 

The main results of this paper are: 
(i) the signatures of the photonic band structure can appear for aggregates as small as 5 particles; (ii) in a cluster of 32 particles the density of (resonant) states can show strong resemblance to that of an infinite structure, with a pronounced depletion in the region of pseudo gap; (iii) photonic band structure of the finite clusters is tolerable to a certain degree of random displacements of the particles off their lattice sites.

We consider  clusters of 5 and 32 spheres, with diameter $0.95$ cm and refractive index $n=3.14$. For both clusters we assumed FCC ordering with the distance between nearest neighbors of 1.9 cm. The system was chosen to match the setup used in the light localization experiment -- Ref. \onlinecite{andrey}. Smaller cluster is comprised of three planar layers perpendicular to the 111 direction, with 1, 3, and 1 sphere in each layer. In the larger cluster, particles were arranged in 5 layers with 3, 7, 12, 7, and 3 spheres in each layer, with the same stacking as in the first cluster. 

The cluster is an open system that does not support stable modes, quasi-states of EM field are leaky, and density of the states cannot be specified. For electrons \cite{fyodorov} in open stochastic systems,  Wigner delay time, $\hat{\tau}_w(E)=\hbar /i\;\partial ln(\hat{\textbf{S}}) /\partial E$, expressed in terms of scattering matrix $\hat{\textbf{S}}$, can be shown to be proportional to the density of resonant modes  
\begin{equation}
\tau_w(\omega )\sim \sum_j\frac{\gamma_j /2}{(\omega -\omega_j)^2+(\gamma_j/2)^2},
\label{wigner}
\end{equation}
where $\omega_j$ and $\gamma_j$ are the energy and the linewidth of the $j$th resonance. For an infinite system, the poles of the scattering matrix $\hat{\textbf{S}}$ approach the real energy axis and the stationary modes are formed, $\gamma_j$ become zero. For EM waves, one can define dwell time, $\tau_d$, that closely follows EM version of the Wigner time, giving asymptotically the same result on the resonances \cite{bart}. Dwell time can be readily found in terms of scattering coefficients as follows:
\begin{equation}
\tau_d(\omega )=\frac{1/N\;\sum_i^N \int_{S_i} W(\textbf{r},\omega )d\textbf{r}}{c\; \sigma_{scat}(\omega )},
\label{eq:dwell}
\end{equation}
where $W(\textbf{r},\omega )$ is electromagnetic energy normalized to 1 in vacuum, $\sigma_{scat}$ is the scattering cross-section, normalized by the total geometrical cross-section $\sum_i^N \pi r_i^2$, and  $c$ is speed of light. The integral in Eq. (\ref{eq:dwell}) is taken over the volume of each scatterer, $S_i$. It gives $\tau_d$ in units of $r/c$ and numerically coincides with cavity quality factor $Q$. The above expression can be understood physically as a ratio between the energy stored in a scatterer divided by outgoing energy current.  $\int_{S_i} W(\textbf{r},\omega )d\textbf{r}$ can be found by introducing an infinitesimally small absorption in the refractive index of the spheres $n+i\kappa$\cite{bart} 
\begin{equation}
\int_{S_i}W(\textbf{r},\omega )\simeq  \frac{3}{8}n\lim_{\kappa \rightarrow 0}\frac{Q_{abs\; i}}{x_i \kappa},
\label{eq:w}
\end{equation}
where $Q_{abs\; i}$ is the absorption cross-section, and $x_i=2\pi r_i/\lambda$ is size parameter of the $i$th sphere. The problem of finding quasi-DOS of the finite cluster is now reduced to finding the scattering coefficients $Q_{abs\; i},\;\sigma_{scat}$. This step can be done by evaluating the expansion series of GMM solution \cite{gmm}. To obtain a measure of the total density of resonant states $\tau_d^{tot}$, expression Eq.(\ref{eq:dwell}) was averaged over 9 symmetrically non-equivalent orientations (altogether 108 directions, accounting symmetry). We would like to stress, that within this framework, the source of error are numerical accuracy of the evaluation of expansion coefficients and limited number of orientations used in angular averaging, both of them can be controlled. 

\begin{figure}
\includegraphics{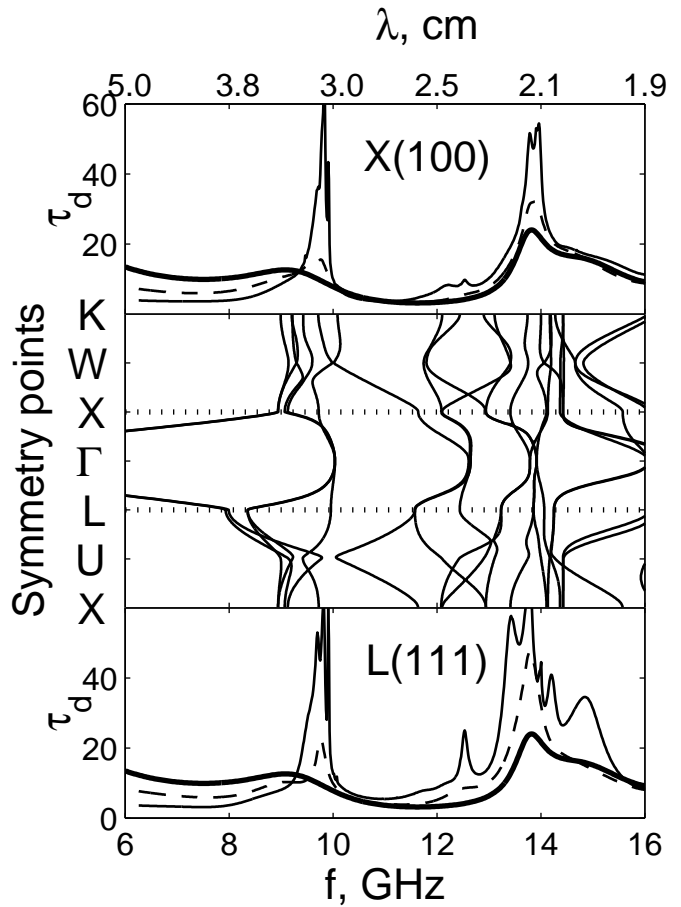}
\caption{\label{disp} Photonic crystal band structure (middle) is compared to the dwell times obtained for scattering from a single sphere (bold solid line), cluster of 5 spheres (dashed line), and the cluster of 32 spheres (thin solid line). Upper graph corresponds to EM plane wave incoming in 100 (X), the lower graph --- 111 (L) crystallographic direction.}
\end{figure}

One characteristic property of a photonic crystal is spacial dispersion, $\omega (\textbf{k})$. On Fig. \ref{disp} we compare dispersion curves of the infinite photonic crystal, obtained using  MIT photonic bands code \cite{mpb}, to the dwell times for two clusters, described above, calculated for 2 different incident angles corresponding to $X$ and $L$ crystallographic directions. It can be easily seen, that even for smaller cluster of 5 particles $\tau_d$ shows the formation of the resonant modes at the positions of photonic modes in $L$-direction. For the cluster of 32 particles, well defined mode structure appears in both directions. However it is more pronounced for $L$ direction, where 6 resonant states between 12 and 16 GHz can be traced to $L$-modes of the infinite photonic crystal. The shift of the modes can ascribed to the coupling between different modes owing to the finite size of the system. Indeed, in the infinite structure the conservation of momentum of EM wave, due to periodicity, would decouple $L$-modes, whereas the finite coupling in truncated crystal leads to the shift and widening of the resonances. Based on symmetry considerations, we can understand the stronger $L$-modes as compared to $X$-modes. The former are formed due to Bragg reflections from 3 or 5 (for the smaller and larger clusters respectively) planes formed by adjacent (nearest) neighbors in 111 direction, while in $X$ direction, Bragg planes are formed only by next-nearest neighbors, which would make them more susceptible to the truncation. 

\begin{figure}
\includegraphics{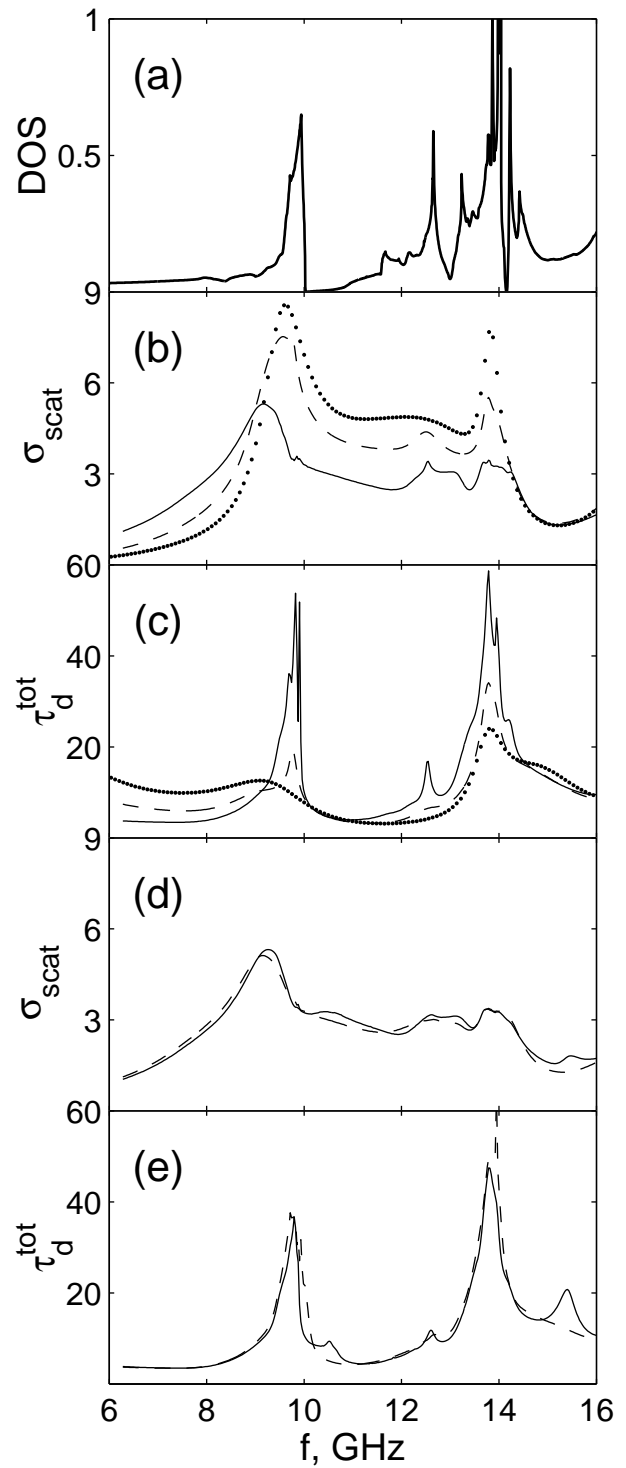}
\caption{\label{scat} (a) Density of EM states in the infinite photonic crystal. (b) Normalized scattering cross-section for a single sphere (dotted), and clusters of 5 (dashed) and 32 (solid) spheres. (c) Normalized dwell time for the same as in (b). (d) Normalized scattering cross-section for cluster of 32 spheres, displaced by  half-radius in random direction off their lattice position (dashed line), and a cluster of 32 spheres, where 5 spheres were replaced by the smaller, 0.85 cm diameter, spheres (solid line). (e) Delay time for the same as in (d).}
\end{figure}

$\tau_d^{tot}$ and angularly (as for $\tau_d^{tot}$) averaged $\sigma_{scat}$ in the region of the first two single sphere Mie resonances are shown on Figs. \ref{scat}(b,c). For comparison we included the total density of EM states, Fig. \ref{scat}(a), calculated for the infinite crystal using the method described in Ref. \onlinecite{busch_john}. Judging from scattering cross-section only, one cannot compare a finite cluster to the infinite structure, while $\tau_d^{tot}$ allows straightforward comparison with DOS of the photonic crystal

Fig. \ref{scat}(b) shows twofold decrease of  $\sigma_{scat}$ at Mie resonances in the cluster. This suppression of the scattering efficiency can be attributed to the hybridization of the single particle resonances \cite{moroz_resphc} due to multiple scattering. Dwell time is closely related to the near-fields \cite{bart} inside the clusters, where formation of the Bragg standing modes leads to substantial modification of the spectrum -- Fig. \ref{scat}(c). For 32-sphere cluster one can see strong resemblance between the density of the resonant modes expressed by $\tau_d^{tot}$ and DOS in the photonic crystal, Fig. \ref{scat}(a). It is interesting to compare our results to 1D case considered in Ref. \onlinecite{mqw}. For a stack of periodically arranged quantum wells it was argued that with increase of the system size, so-called, subradiant EM modes formed stable modes of the photonic passband with large life-times, similar to our result for dwell times in the 3D clusters.

To assess the sensitivity of our results to disorder in the cluster, we performed the calculations of $\sigma_{scat}$ and $\tau_d^{tot}$, Fig \ref{scat}(d,e), for: the cluster of 32 spheres displaced by the half  of the particle radius in random directions (dashed lines) off their initial positions and the cluster of 32 spheres where 5 spheres replaced by the defect spheres of a different diameter (0.85 cm) - solid lines. One can see, that two types of disorder had different effect on the spectra. While positional (topological) disorder led only to smearing of only sharp resonances, the defects also introduced a number of new peaks. The later effect can be related to the new resonances introduced by the defect spheres. The effect of the topological disorder on photonic bandgaps was studied in (finite but comprised of large number of particles) 2D \cite{soukdis2d,asatryan} and 3D photonic crystals \cite{zhang}. Our conclusion on stability of the band structure to the topological disorder seems to be in line with Ref. \onlinecite{soukdis2d}. However, besides obvious difference in the dimensionality of the space (we work in 3D), another important difference exists. In large systems, to explain the existence of the band structure one can resort to the periodic-on-average argument, while in such a small system as ours one cannot.

It is worth noting, that FCC opal structures are not usually considered as a good candidate for potential applications connected to complete photonic bandgap \cite{opal_vector}. However, as it can be seen from our example, for some parameters, DOS can be significantly suppressed in a wide spectral region. Fig. \ref{scat} also suggests that such structures can be tolerable to strong topological disorder.

As it was already mentioned in the introduction, our results are relevant for the problem of light propagation in a dense random media made of particles with a narrow size dispersion. Topological disorder does not prevent the occurrence of ordered clusters. In Ref. \onlinecite{andrey}, random displacement of scatterers off their lattice positions did not have a significant effect on the density of EM states. Our calculations performed for the clusters with similar random displacements (Fig. \ref{scat}(d,e)), confirm  small effect on the formation of the photonic band structure. This also suggests that the light localization observed in Ref. \onlinecite{andrey} may have been facilitated by depletion of the density of electromagnetic states due to the presence of the pseudogap. 

It can be further argued that when the scattering length $l_{scat}$ becomes of the same order as the mean distance between the particles, the occurrence of clusters cannot be ignored. Indeed, the scattering inside the clusters is dominated by Bragg reflection mechanism, while $l_{Bragg}$ can be as small as a few lattice constants for high index contrast structures. Furthermore, in Ref. \cite{cbs_opal} it was shown that periodicity of a photonic crystal may also exhibit itself in coherent back-scattering effect. The modification of coherent back-scattering due to the presence of ordered clusters is an interesting problem that deserves a detailed consideration.

In conclusion, we calculated scattering cross-section and density of resonant states in small 3D ordered aggregates of dielectric spheres within framework of the rigorous GMM solution. In contrast with previous studies we made no assumptions about the strength, size, or separation between scatterers.  All multiple scattering effects were automatically retained in the solution. The results suggest that the photonic band structure of the infinite crystal can show up for clusters as small as 5 particles. Density of resonant states is significantly perturbed compared to single particle case. For a cluster of 32 particle the density of resonant states can show a close resemblance to DOS of the infinite system.

\begin{acknowledgments}
AY gratefully acknowledges numerous insightful discussions and comments by A. A. Chabanov, A. A. Lisyansky, and L. I. Deych. This work is supported by the National Science Foundation under Grant No DMR-0093949. HC acknowledges the support from the David and Lucile Packard Foundation and Alfred P. Sloan Foundation.
\end{acknowledgments}

\bibliography{clusters}

\end{document}